\documentclass[3p,shortbib,twocolumn]{elsarticle}

\usepackage{amsmath,amssymb}
\usepackage{graphicx,epstopdf}
\usepackage{graphicx}
\usepackage[usenames,dvipsnames]{xcolor}
\usepackage{slashed}
\usepackage{ulem}
\usepackage[dvipsnames]{xcolor}

\usepackage{hyperref}
\hypersetup{colorlinks=true}

\newcommand{\be}{\begin{equation}}
\newcommand{\ee}{\end{equation}}
\newcommand{\bea}{\begin{eqnarray}}
\newcommand{\eea}{\end{eqnarray}}

\definecolor{aaaa}{rgb}{0.99, 0.4, 0.01}
\definecolor{bbbb}{rgb}{0.5, 0.3, 0.9}

\newcommand{\bi}{\begin{itemize}}
\newcommand{\ei}{\end{itemize}}

\makeatletter
\def\ps@pprintTitle{%
 \let\@oddhead\@empty
 \let\@evenhead\@empty
 \def\@oddfoot{}
 \let\@evenfoot\@oddfoot}
\makeatother

\newcommand{\lambdabar}{{\mkern0.75mu\mathchar '26\mkern -9.75mu\lambda}}

\begin{document}

\begin{frontmatter}

\title{Diffusion as a possible mechanism controlling the production of superheavy nuclei in cold fusion reactions}
\author{T.~Cap}
\author{M.~Kowal\corref{Cof}}
\cortext[Cof]{Corresponding author}
\ead{michal.kowal@ncbj.gov.pl}
\address{National Centre for Nuclear Research, Pasteura 7, 02-093 Warsaw, Poland}
\author{K.~Siwek-Wilczyńska}
\address{Faculty of Physics, Warsaw University, Pasteura 5, 02-093 Warsaw, Poland}

\begin{abstract}
The fusion probability for the production of superheavy nuclei in cold fusion reactions was investigated and compared with recent experimental results for $^{48}$Ca, $^{50}$Ti, and $^{54}$Cr incident on a $^{208}$Pb target. Calculations were performed within the fusion-by-diffusion model (FbD) using new nuclear data tables by Jachimowicz et al.
It is shown that the experimental data could be well explained within the framework of the FbD model. The saturation of the fusion probability at bombarding energies above the interaction barrier is reproduced.  It emerges naturally from the physical effect of the suppression of contributions of higher partial waves in fusion reactions and is related to the critical angular momentum. The role of the difference in values of the rotational energies in the fusion saddle point and contact (sticking) configuration of the projectile-target system is discussed.
\end{abstract}

\end{frontmatter}

\section{Introduction}

Superheavy elements with atomic numbers $104 \le Z \le 113$ were discovered in cold fusion reactions in which closed-shell $^{208}$Pb or $^{209}$Bi target nuclei were bombarded with projectiles ranging from Ti to Zn~\cite{Hofmann,Riken}.

The production cross section for superheavy nuclei (SHN) can be considered as the product of three factors: the cross section for the projectile to overcome the entrance channel barrier (capture cross section), the probability that the resulting system fuses and reaches the compound nucleus configuration, and the probability, that the excited compound nucleus survives fission during deexcitation.

The optimal conditions for obtaining a given superheavy nucleus result from various factors. The increase in the symmetry between reaction partners requires higher bombarding energies to overcome the entrance channel barrier and enhances the contribution of fast non-equilibrium deep-inelastic (DIC) and quasi fission (QF) processes preventing fusion. The fusion probability rapidly drops when the product of projectile and target nuclei atomic numbers $Z_{1}\times Z_{2} \geq 1600 $ ~\cite{Schmidt-91}. Therefore, the compound nucleus formation cross section represents only part of the capture cross section.

In cold fusion reactions, merging the strongly bound target and projectile nuclei leads to a weakly bound compound nucleus. Typically, SHN have higher thresholds for neutron emission~\cite{Jach2021} than the heights of the fission barrier~\cite{Jach2017}, making fission the dominant deexcitation process. At each step of the deexcitation cascade, neutron evaporation competes with fission, which additionally reduces the final evaporation residue cross section.

The cross sections for the production of nuclei with $Z\ge102$ in cold fusion reactions drop approximately seven orders of magnitude as the projectile atomic number changes from 20 (Ca) to 30 (Zn). The question what is the mechanism preventing the synthesis of SHN is still under discussion. A low survival probability is not enough to explain the extremely low production cross sections. One way of thinking about the additional hindrance mechanism is the concept of an internal barrier holding back (counteracting) the fusion process. Overcoming this barrier by a diffusion process and thermal fluctuations could help to reach the state of a compound nucleus.

Recently, the probability of compound nucleus formation $P_{CN}$, at energies around and above the interaction barrier $B_0$ was measured for $^{48}$Ca, $^{50}$Ti, and $^{54}$Cr projectile incident on a $^{208}$Pb target~\cite{HindePRL}. The authors of Ref.~\cite{HindePRL} concluded that ``the energy dependence of $P_{CN}$ indicates that cold fusion reactions (involving $^{208}$Pb) are not driven by a diffusion process''. This letter aims to investigate the fusion probabilities for these reactions using the diffusion approach. Calculations are performed within the $l$-dependent fusion-by-diffusion model (FbD)~\cite{FBD-11} using the new nuclear data tables for SHN by Jachimowicz et al.~\cite{Jach2021} as input.

\section{FbD model}~\label{SECT:THEORY}

The fusion-by-diffusion model in its first form was a simple tool to calculate cross sections and optimum bombarding energies for a class of $1n$ cold fusion reactions~\cite{FBD-Acta,FBD-05}. A significant development of this model was the incorporation of the angular momentum dependence, that is, the contributions from successive partial waves to the reaction cross section~\cite{FBD-11}.

Due to the different time scales of the particular reaction stages, the partial evaporation residue cross section, $\sigma_{ER}(l)$, can be factorized as the product of the partial capture cross section $\sigma_{cap}(l) = \pi \lambdabar^2(2l + 1)T(l)$, the fusion probability $P_{fus}(l)$, and the survival probability $P_{surv}(l)$. Thus, the total evaporation residue cross section for the production of a given superheavy nucleus in its ground state is
\begin{equation}
\label{factorize}
 \sigma_{ER} = \pi \lambdabar^2 \sum_{l = 0}^{\infty}(2l+1)T(l)\times P_{fus}(l)\times P_{surv}(l),
\end{equation}
where $\lambdabar$ is the wavelength, and $\lambdabar^2=\hbar^2/2\mu E_{c.m.}$. Here $\mu$ is the reduced mass of the colliding system, and $E_{c.m.}$ is the center-of-mass energy at which the reaction takes place.

The method of calculating the capture cross section is described in the next section. The fusion probability is described in detail in section~\ref{SOBSECT:FUS}.

The last factor in Eq.~\ref{factorize}, the survival probability, is calculated by applying classical transition state theory using nuclear data from Ref.~\cite{Jach2021}. Details regarding this reaction stage for $1n$ cold fusion reactions can be found in Ref. \cite{FBD-11}.

\subsection{Capture cross section}~\label{SOBSECT:CAP}
The capture transmission coefficients $T(l)$ in Eq. \ref{factorize} are calculated in a simple sharp cut off approximation, where the upper limit $l_{max}$ of full transmission, $T(l)=1$, is determined from the empirical systematics of the capture cross sections for heavy nuclear systems.

Following the experimental results, the entrance channel barrier is not described by a single value but by a distribution that can be approximated by a Gaussian shape described by two parameters, the mean barrier $B_{0}$ and the distribution width $\omega$~\cite{KSW04}. Folding the Gaussian barrier distribution with the classical expression for the fusion cross section leads to the formula for the capture cross section
\begin{eqnarray}
\label{capture}
\sigma_{cap} &=& \pi R^{2} \frac{\omega}{E_{c.m.} \sqrt {2\pi}} \Big[X\sqrt{\pi} (1 + \textrm{erf}(X))\nonumber\\&+&\exp(-X^{2})\Big] = \pi \lambdabar^2 (2l_{max}+1)^2,
\end{eqnarray}
where $X = \frac{E_{c.m.}-B_{0}}{\omega \sqrt{2}}.$
The empirical systematics of $B_0$, $\omega$, and the normalization factor $R$ were obtained from analyzing precisely measured fusion or capture excitation functions for about 50 heavy nuclear systems for which the fusion probability is equal or close to unity~\cite{KSW04}. In this paper we use the parametrizations of $B_0$, $w$, and $R$ of Ref.~\cite{FBD-11}.

\subsection{Fusion probability}~\label{SOBSECT:FUS}

The second factor in Eq.~\ref{factorize}, $P_{fus}(l)$, is the probability that after reaching the capture configuration, the colliding system will eventually overcome the fusion saddle point and merge, avoiding reseparation. It is assumed in the FbD model that after sticking, a neck between the target and projectile nuclei rapidly grows at an approximately fixed mass asymmetry and elongation \cite{FBD-Acta, FBD-05} bringing the system to the ``injection point'' somewhere along the bottom of the asymmetric fission valley. Let us denote the elengation of the system at the ``injection point'' by $L_{inj}$. The localization of this point with respect to the macroscopic conditional saddle (at the elongation $L_{sd}$) is crucial for the fusion process. If $L_{inj} > L_{sd}$ the system is still ``outside" the barrier separating the ``injection point'' from the compound nucleus configuration and must climb uphill to overcome the saddle. If $L_{inj} < L_{sd}$ the ``injection point'' configuration is more compact than the saddle configuration, and the system is already ``inside" (behind the barrier). In this case, the barrier guards the system against reseparation by reducing the outgoing flux of particles.

In the diffusion approach, transition over the barrier happens by thermal fluctuations in the shape degrees of freedom. The fusion probability, $P_{fus}(l)$, may be derived by solving the Smoluchowski diffusion equation. With the assumption that the internal barrier has height $H(l)$ and is of inverted parabola form one gets
\cite{FBD-Acta}
\begin{align}\label{pfus}
  P_{fus}(l) = \frac{1}{2} \left\{
\begin{array}{lr}
    1 + { \rm erf}\sqrt{ H(l)/T} & : L_{inj} < L_{sd} \\
    1 - { \rm erf}\sqrt{ H(l)/T} & : L_{inj} \ge L_{sd} \\
  \end{array},
\right.
\end{align}
where $T$ is the average temperature of the fusing system (see~\cite{FBD-11} for details).

The energy threshold in Eq.~\ref{pfus} is taken as the difference between the energy of the fusion saddle point $E_{sd}$ and the energy of the combined system at the ``injection point'' $E_{inj}$, corrected by the appropriate rotational energies,
\begin{equation}\label{H}
H(l) = (E_{sd} + E_{sd}^{rot}) - (E_{inj} + E_{inj}^{rot}).
\end{equation}

The energies $E_{sd}$ and $E_{inj}$ are calculated using simple algebraic expressions that approximate the potential energy surface \cite{FBD-11}. The shape parametrization used to describe the interacting system is that of two spheres joined smoothly by a third quadratic surface. The corresponding values of the rotational energies at the injection point $E_{inj}^{rot}$ and the saddle point $E_{sd}^{rot}$ are calculated assuming the rigid-body moments of inertia for the respective shapes~\cite{FBD-11}.

The distance between the nuclear surfaces of two colliding nuclei at the injection point, $s_{inj}$, is
the only adjustable parameter of the model. It defines the onset of the diffusion process, thus, the moment when the available kinetic energy that remains after passing the entrance barrier is already transformed into internal degrees of freedom in the over-damped regime.

\begin{figure}[t!!]
 \center{\includegraphics[width=0.8\columnwidth,angle=-90]{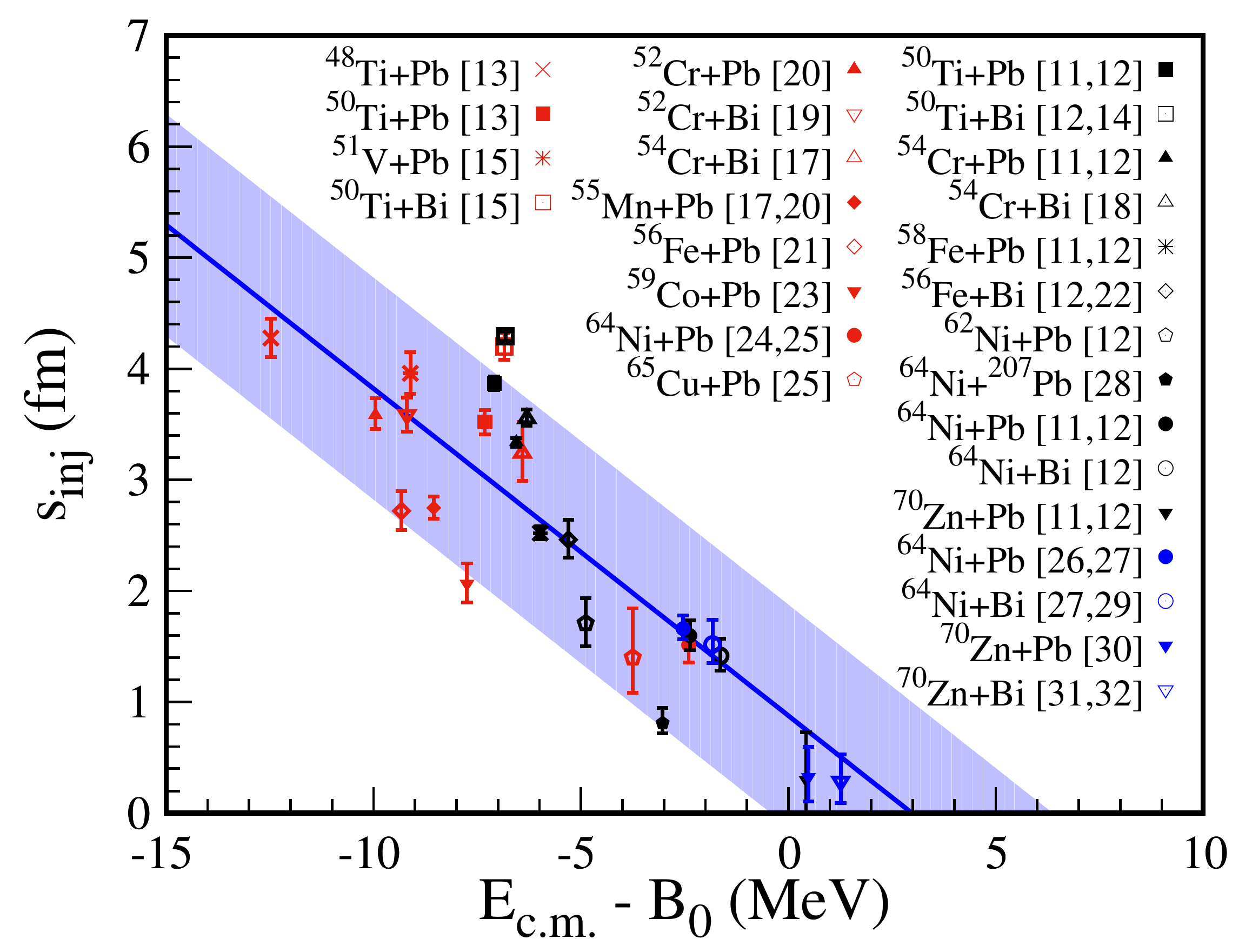}}
\caption{\label{Fig_sinj} The injection point systematics obtained for the set of $1n$ cold fusion reactions \cite{HOFMANN200493,Hofmann_1998,PhysRevC.78.024605,ISI:000172568800008,PhysRevC.78.034604,PhysRevC.79.027602,PhysRevC.78.024606,ISI:A1989U697600008,PhysRevLett.100.022501,PhysRevC.73.014611,PhysRevC.79.011602,ISI:A1997XX26900004,PhysRevC.79.027605,PhysRevC.67.064609,PhysRevLett.93.212702,ISI:000223717300009,MORITA2004101,ISI:000167664500002,doi:10.1143/JPSJ.73.1738,doi:10.1143/JPSJ.76.043201,doi:10.1143/JPSJ.73.2593,New113} using new nuclear data tables~\cite{Jach2021}. If not indicated otherwise, targets were $^{208}$Pb or $^{209}$Bi. The color of the points indicates the laboratory where the reaction was studied: LBNL (red), GSI (black), RIKEN (blue). See text for details. }
\end{figure}

In this paper, we redefine the systematics of this crucial parameter based on a new set of ground state and saddle point properties of SHN \cite{Jach2021}. The new parametrization of the ``injection point distance'' is shown in Fig.~\ref{Fig_sinj} as a function of the excess of the center-of-mass energy $E_{c.m.}$ over the mean barrier $B_{0}$. Each point represents the value of the $s_{inj}$ distance obtained by fitting Eq.~\ref{factorize} to the experimentally measured $1n$ evaporation residue cross sections for 27 cold fusion reactions (see Ref.~\cite{FBD-11} for fitting protocol details).

It can be seen from Fig.~\ref{Fig_sinj} that for energies up to a few MeV above $B_{0}$ the $s_{inj}$ distance can be well approximated by a straight line given by
\begin{equation}\label{sinj}
s_{inj} = 0.878~\textrm{fm} - 0.294\times(E_{c.m.} - B_{0})~\textrm{fm/MeV}.
\end{equation}
A similar linear trend of $s_{inj}$ as a function of $E_{c.m.} - B_{0}$ was reported in Ref.~\cite{Boilley} by solving Langevin type equations.

The shaded area in Fig.~\ref{Fig_sinj} represents an error corridor of $\pm 1$ fm, which allows the uncertainty of the calculated fusion probabilities to be determined. The parametrization given by Eq.~\ref{sinj} should be used for interpolation rather than extrapolation far beyond the explored range of $ E_{c.m.} - B_{0}$ values, especially if the extrapolation leads below the physically acceptable limit of the touching configuration $(s_{inj} \approx 0)$. Negative values of this parameter would correspond to a large overlap of the density distributions at the sticking stage, an effect that is impossible in nuclear collisions at low kinetic energies. Therefore, in collisions at energies higher than a few MeV above $B_{0}$, we assume $s_{inj}=0$ (allowing a deviation in the range of 1 fm).

\section{Results and discussion}~\label{SECT:RES}

\begin{figure}[t!!]
\center{\includegraphics[width=8cm,height=7cm,angle=0.0]{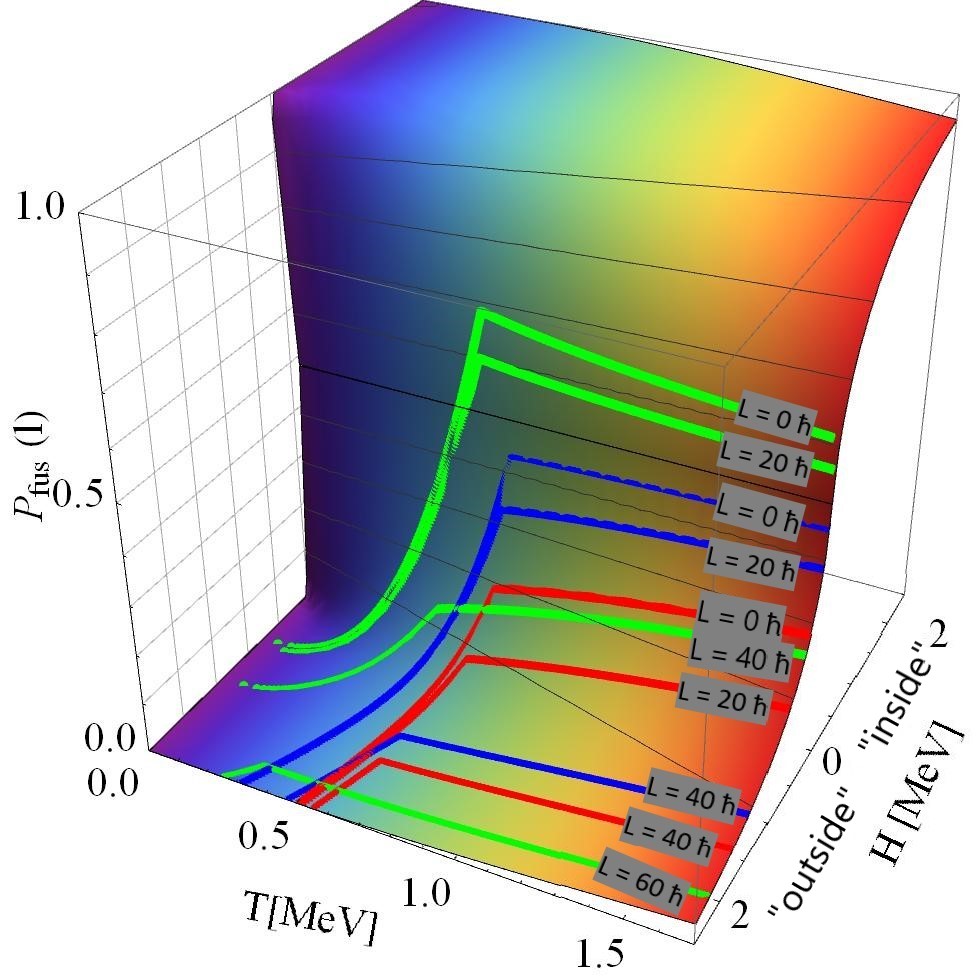}}
\caption{\label{3d} Fusion probability $P_{fus}(l)$ as a function of temperature $T$ and the barrier height  opposing fusion $H(l)$ for angular momenta $l = 0 \hbar, 20 \hbar, 40 \hbar$, and $60 \hbar$. Calculations for $^{48}$Ca +$^{208}$Pb (green lines), $^{50}$Ti +$^{208}$Pb (blue lines) and $^{54}$Ca +$^{208}$Pb (red lines) fusion reactions. The color of the surface marks the temperature gradient of the synthesized system. Labels ``outside" and ``inside" refer to the ``injection point'' position to the saddle (see Eq.~\ref{pfus} and its discussion).}
\end{figure}

The FbD model with the new ``injecton point distance'' parameterization and input data taken from Ref.~\cite{Jach2021} was used to analyze fusion probabilities for $^{48}$Ca, $^{50}$Ti, and $^{54}$Cr reactions incident on a $^{208}$Pb target.

The fusion probability given by Eq.~\ref{pfus} depends on the ratio of the height of the barrier opposing fusion $H(l)$ and the average temperature $T$ of the system during the merging process. The fusion probabilities, $P_{fus}(l)$, for $^{48}$Ca +$^{208}$Pb  (green lines),  $^{50}$Ti +$^{208}$Pb (blue lines), and $^{54}$Ca +$^{208}$Pb  (red lines) reactions as a functions of $H(l)$ and $T$ are shown in Fig.~\ref{3d} for a few selected $l$ values.

Let us start the discussion by analyzing the $l=0$ case in which the height of the barrier is simply the difference between the asymmetric conditional saddle point energy and the energy of the combined system of the projectile and target nuclei separated by the distance $s_{inj}$ (see Eq.~\ref{H}). As the available energy increases, the ``injection point distance'' decreases (see Fig.~\ref{Fig_sinj}), leading to a lowering of the barrier height, and thus to the rapid growth of the fusion probability (see  Fig. 2). When the separation distance reaches zero at the energy corresponding to $T \approx 0.6 -0.8$ MeV, the fusion probability reaches the maximum in all three analyzed reactions. Further energy increase does not change the height of the barrier ($s_{inj}$ remains equal to zero) but heats the system up and thus affects the fusion probability. For $^{48}$Ca +$^{208}$Pb reaction, the touching configuration is behind the asymmetric saddle point (as seen from the entrance channel perspective; ``inside" regime in Fig.~\ref{3d}). In this case, the rising temperature increases the flux of particles escaping through the asymmetric saddle point and thus slightly reduces the fusion probability (see $L_{inj} < L_{sd}$ case in Eq.~\ref{pfus}). For $^{50}$Ti and $^{54}$Cr projectiles, the touching configuration is ``outside" the barrier and the fusion probability slowly increases with the increase of the incindent energy ($L_{inj} \ge L_{sd}$ case in the Eq.~\ref{pfus}).

The inclusion of the higher partial waves affects the entire potential energy surface topology and influences the competition between the existing symmetric and asymmetric saddle points. In particular, the symmetric saddle, being more compact and having a lower moment of inertia, is more sensitive to the increase of the angular momenta. Above a certain $l$-value, the symmetric saddle begins to dominate and becomes the main point to overcome in the fusion process for all studied systems. In this case, the solution of the Smoluchowski diffusion equation can also be applied to calculate the fusion probability. However, the barrier height $H(l)$ should be calculated with respect to the symmetric saddle point. The barrier height increases with the increase of the $l$-value due to the difference in the rotational energies in the symmetric saddle and ``injection'' points. Therefore, the contribution of higher partial waves to the fusion cross section is suppressed. The systematic decrease of the fusion probability with the increase of the $l$-value observed in Fig.~\ref{3d} for all three reactions might be viewed as a manifestation of the well-known effect of the critical angular momentum.

As one can see in Fig.~\ref{3d}, the dominant contribution to the analyzed cold fusion reactions comes from near-central collisions. The more peripheral collisions are less favorable and lead to the re-separation of the system at the beginning of the nuclear reaction, rather than merging of target and projectile nuclei.
\begin{figure*}[t!!]
\center{\includegraphics[width=1.1\columnwidth,angle=-90]{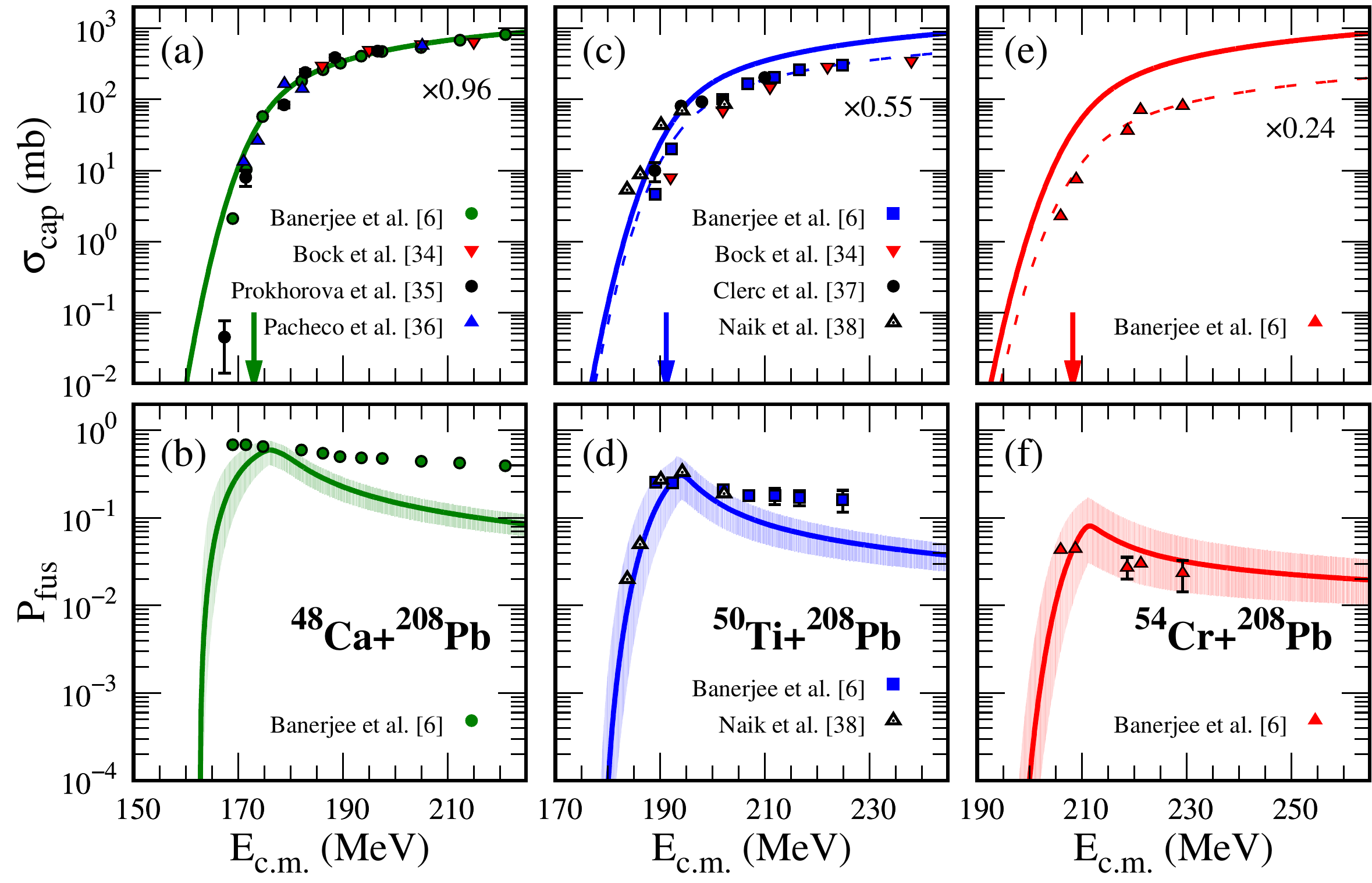}}
\caption{\label{multiplot} Capture cross section $\sigma_{cap}$ (top panels) and averaged  fusion probability $P_{fus}$ (bottom panels) for the reactions: $^{48}$Ca+$^{208}$Pb - panels (a), (b), $^{50}$Ti +$^{208}$Pb -  panels (c), (d), $^{54}$Cr +$^{208}$Pb -  panels (e), (f). Solid lines show the FbD model calculations of $\sigma_{cap}$ and $P_{fus}$. Dashed lines in the top panels show calculated $\sigma_{cap}$ scaled by the indicated suppression factors. The arrows indicate the value of the mean entrance channel barrier, $B_{0}$, for each reaction. The error corridors resulting from the $s_{inj}$ systematics uncertainty are shown as shaded areas in the bottom panels. Points represent relevant experimental data taken from Refs.~\cite{HindePRL,BOCK,PROKH,Pacheco,CLERC,Naik}. If not shown, error bars are smaller than the symbol sizes.}
\end{figure*}

To study the effective fusion probability for a given reaction, one can define the quantity
\begin{equation}
\label{fuseq}
 P_{fus} =\frac {1}{(2l_{max}+1)^2}\sum_{l = 0}^{l_{max}}(2l+1)\times P_{fus}(l),
\end{equation}
which is the fusion probability ``averaged'' over all angular momenta contributing to the fusion cross section.

In Fig.~\ref{multiplot} we present a comparison of the FbD model predictions with the experimental data. The top panels show the capture cross sections for each of the reactions (i.e. cross sections for overcoming the entrance channel barrier) calculated using Eq.~\ref{capture}. Model calculations are compared with experimental data taken from Ref.~\cite{HindePRL,BOCK,PROKH,Pacheco,CLERC,Naik}. The arrows in panels (a), (c), and (e) indicate the values of the mean entrance channel barriers $B_0$, calculated using the empirical parametrization~\cite{FBD-11}, $173.0$, $191.2$, and $208.3$ MeV for $^{48}$Ca, $^{50}$Ti, and $^{54}$Cr respectively.

The experimentally measured fission-like cross sections shown in the top panels of Fig.~\ref{multiplot} lie below our calculations (solid lines). The deviation increases with increasing projectile atomic number. As proposed in \cite{HindePRL}, we estimated scaling factors, $S$, for our calculations to reproduce the experimental results in the energy range above $B_0$. These factors are 0.96 for $^{48}$Ca+$^{208}$Pb, 0.55 for $^{50}$Ti+$^{208}$Pb and 0.24 for $^{54}$Cr+$^{208}$Pb (in \cite{HindePRL} the respective factors are 0.75, 0.48 and 0.22). Scaled capture cross sections are shown as dashed lines in Fig.~\ref{multiplot}.

Our scaling factors are in reasonable agreement with the results presented in Ref.~\cite{HindePRL}, where they were estimated as a deviation from the CCFULL model based on the coupled channels formalism \cite{Hagino}. The capture cross section suppression might be associated with mass-asymmetric fast non-equilibrium processes, such as QF or DIC, appearing just after the interacting system passes the entrance channel barrier. It should be emphasized that both in this work and Ref.~\cite{HindePRL}, the obtained scaling factors are model-dependent.

Calculated average fusion probabilities (see Eq.~\ref{fuseq}) for $^{48}$Ca, $^{50}$Ti, and $^{54}$Cr reactions on a $^{208}$Pb target are shown in the lower panels of Fig.~\ref{multiplot}. Full points in panels (b), (d), and (f) represent upper limits on the compound nucleus formation probabilities $P_{sym}$ taken from Ref.~\cite{HindePRL}. $P_{sym}$ is derived as the ratio of the measured symmetric-peaked fission cross section $\sigma_{sym}$ to the capture cross section taken as the measured total fission-like cross section $\sigma_{fis}$ divided by the appropriate scaling factor $S$ ($P_{sym} = \frac{\sigma_{sym}}{\sigma_{fis}/S}$, see Eq. 1 in the supplementary material for~\cite{HindePRL}).

For the $^{50}$Ti+$^{208}$Pb reaction additional experimental points (open triangles in Fig.~\ref{multiplot}(d)) taken from \cite{Naik} are shown. These data were derived by measuring the angular distribution of mass-symmetric fission.

The calculated average fusion probabilities (Eq.~\ref{fuseq}, solid lines in panels (b),(d), and (f)) are in good agreement with the experimental data for all studied reactions. A rapid decrease of the fusion probability in the energy region below $B_0$ reported in~\cite{Naik} for the $^{50}$Ti+$^{208}$Pb reaction is reproduced in our calculations (see Fig.~\ref{multiplot}(d)). Unfortunately, the data for other reactions are limited in this energy region.

For each reaction, the maximum value of $P_{fus}$ is reached for an energy a few MeV above $B_{0}$ (when $s_{inj} \approx 0$). Thus, the steady decrease of $P_{fus}$ with increasing energy is due to the dependence of $H(l)$ on the angular momentum only (see Fig.~\ref{3d}).

\begin{figure}[t!!]
 \center{\includegraphics[width=1.0\columnwidth,angle=-90]{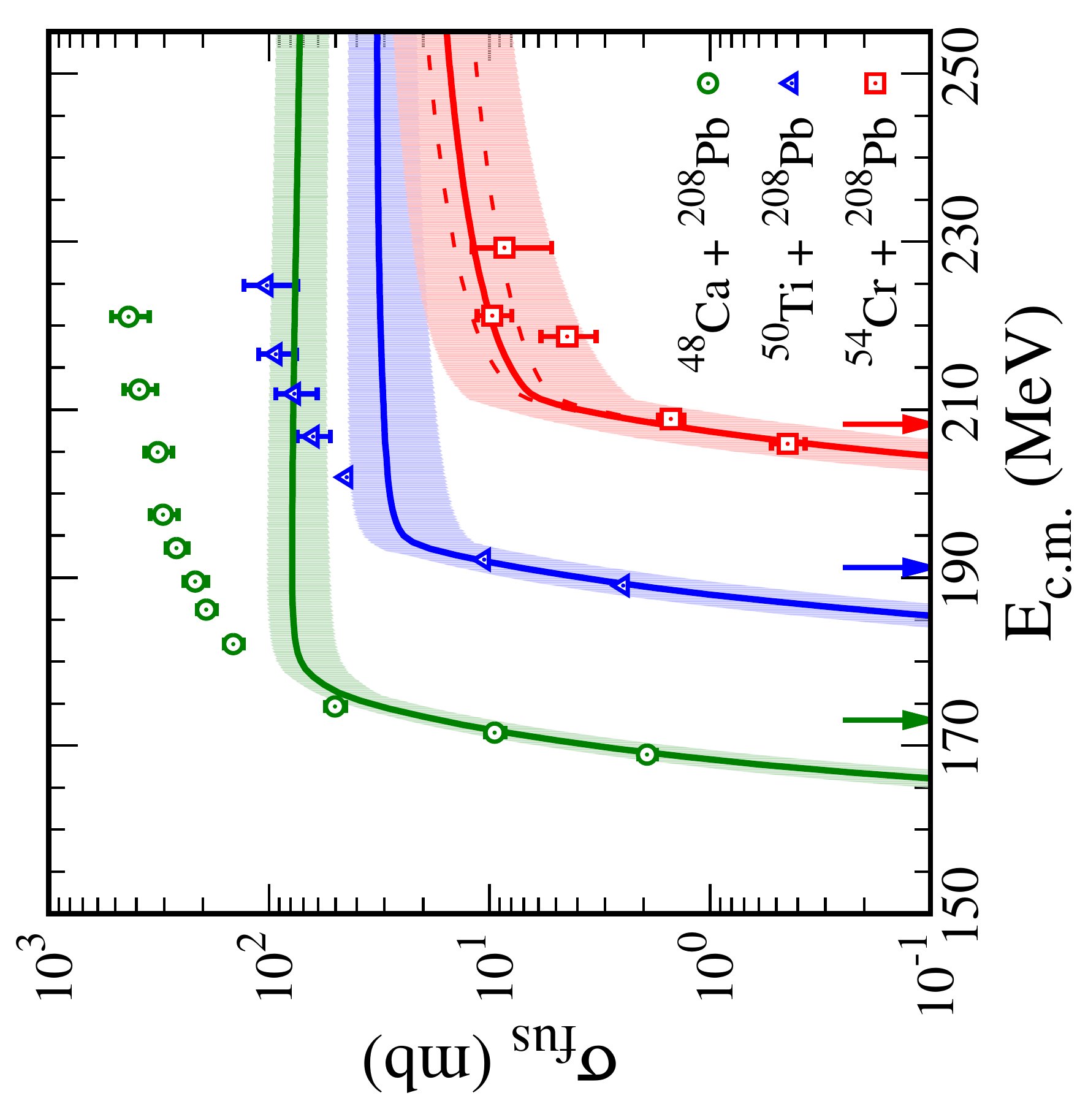}}
\caption{\label{PCN} Calculated compound nucleus formation cross sections, $\sigma_{fus}$, for $^{48}$Ca +$^{208}$Pb, $^{50}$Ti +$^{208}$Pb, and $^{54}$Ca +$^{208}$Pb  fusion reactions. Points are derived from the experimental data presented in Ref.~\cite{HindePRL}. Dashed lines show calculations for two extreme orientations of target and $^{54}$Cr projectile in the entrance channel: tip to tip (bottom line) and body to body (upper line). The arrows indicate the value of the mean entrance channel barrier, $B_{0}$, for each reaction.
See text for details.}
\end{figure}

Finally, in Fig.~\ref{PCN}, we show the FbD model calculations of compound nucleus formation cross sections, defined as
\begin{equation}
\label{PCNeq}
 \sigma_{fus} = \pi \lambdabar^2 \sum_{l = 0}^{l_{max}}(2l+1)
 T(l)P_{fus}(l) = \sigma_{cap}\times P_{fus}.
\end{equation}
The model calculations (solid lines) are compared with the symmetric-peaked fission cross sections measured in~\cite{HindePRL}, but not given in that paper. Therefore, we deduced $\sigma_{sym}$ values from the data using the relation $\sigma_{sym} = P_{sym} \times (\sigma_{fis}/S)$ (see Fig. 4 in Ref.~\cite{HindePRL} and Fig. 5 in the corresponding supplementary material). Although the comparison of $\sigma_{sym}$ with $\sigma_{fus}$ is not entirely unequivocal, it seems adequate. Qualitatively, one can see very similar behavior of the calculated and experimental cross sections as a function of the center-of-mass energy.
Some additional subtle effects related to the deformation of the $^{54}$Cr projectile might be expected. However, as we have checked (by analyzing collisions at the tip to tip and equatorial configurations; dashed lines in Fig.~\ref{PCN}), the results for extreme orientations in the entrance channel are within the error corridor resulting from the systematics of $s_{inj}$.

\section{Conclusions}~\label{SECT:CONCS}

The presented results show that the compound nucleus formation cross sections and related fusion probabilities for $^{48}$Ca, $^{50}$Ti, and  $^{54}$Cr incident on a $^{208}$Pb target can be well reproduced within the framework of the FbD model.

The experimentally observed dependence of the fusion probability on the energy can be reproduced using the diffusion approach. In the energy range below $B_0$, the fusion probability growth comes from the reduction in the height of the internal barrier opposing fusion with increasing bombarding energy. The fusion probability saturation above $B_0$ results from suppression of the contributions from higher partial waves and can be linked to the critical angular momentum. The difference between rotational energies in the fusion saddle and the contact (sticking) configuration at the beginning of the fusion process plays a major role in compound nucleus formation at energies above $B_0$.

\section*{ACKNOWLEDGEMENTS}
M.K. was co-financed by the National Science Centre under Contract No. UMO-2013/08/M/ST2/00257  (LEA COPIGAL).

\bibliography{fbd_plb}

\end{document}